\documentclass[final]{aipproc}
\usepackage{natbib}
\usepackage{amssymb}

\newcommand{\unitspace}{\ensuremath{\,}}
\newcommand{\usp}{\unitspace}
\newcommand{\numberspace}{\ensuremath{\;}}
\newcommand{\nsp}{\numberspace}
\newcommand{\unitstyle}[1]{\ensuremath{\mathrm{#1}}}
\newcommand{\power}[2]{\ensuremath{{#1}^{#2}}}

\newcommand{\kilo}{\unitstyle{k}}
\newcommand{\Mega}{\unitstyle{M}}

\newcommand{\cm}{\unitstyle{cm}}
\newcommand{\gram}{\unitstyle{g}}

\newcommand{\meter}{\unitstyle{m}}

\newcommand{\second}{\unitstyle{s}}

\newcommand{\Kelvin}{\unitstyle{K}}
\newcommand{\K}{\Kelvin}  

\newcommand{\grampercc}{\gram\usp\power{\cm}{-3}} 
\newcommand{\grampersquarecm}{\gram\usp\power{\cm}{-2}} 

\newcommand{\GramPerSc}{\grampersquarecm}
\newcommand{\erg}{\unitstyle{ergs}} 

\newcommand{\ergspersecond}{\erg\unitspace\power{\second}{-1}}

\newcommand{\eV}{\unitstyle{eV}}        
\newcommand{\MeV}{\Mega\eV} 

\newcommand{\Msun}{\ensuremath{M_\odot}}

\newcommand{\yr}{\unitstyle{yr}}        

\newcommand{\mb}{\ensuremath{m_\mathrm{u}}} 

\newcommand{\satellite}[1]{\emph{#1}}

\newcommand{\beppo}{\satellite{BeppoSAX}}

\newcommand{\code}[1]{\textsc{#1}}

\newcommand{\nonsmoker}{\code{non-smoker}}

\def\refer@jnl#1{{\rmfamily#1}}%
\newcommand\aj{\refer@jnl{AJ}}%
\newcommand\araa{\refer@jnl{ARA\&A}}%
\newcommand\apj{\refer@jnl{ApJ}}%
\newcommand\apjl{\refer@jnl{ApJ}}%
\newcommand\apjs{\refer@jnl{ApJS}}%
\newcommand\ao{\refer@jnl{Appl.~Opt.}}%
\newcommand\apss{\refer@jnl{Ap\&SS}}%
\newcommand\aap{\refer@jnl{A\&A}}%
\newcommand\aapr{\refer@jnl{A\&A~Rev.}}%
\newcommand\aaps{\refer@jnl{A\&AS}}%
\newcommand\azh{\refer@jnl{AZh}}%
\newcommand\baas{\refer@jnl{BAAS}}%
\newcommand\jrasc{\refer@jnl{JRASC}}%
\newcommand\memras{\refer@jnl{MmRAS}}%
\newcommand\mnras{\refer@jnl{MNRAS}}%
\newcommand\pra{\refer@jnl{Phys.~Rev.~A}}%
\newcommand\prb{\refer@jnl{Phys.~Rev.~B}}%
\newcommand\prc{\refer@jnl{Phys.~Rev.~C}}%
\newcommand\prd{\refer@jnl{Phys.~Rev.~D}}%
\newcommand\pre{\refer@jnl{Phys.~Rev.~E}}%
\newcommand\prl{\refer@jnl{Phys.~Rev.~Lett.}}%
\newcommand\pasp{\refer@jnl{PASP}}%
\newcommand\pasj{\refer@jnl{PASJ}}%
\newcommand\qjras{\refer@jnl{QJRAS}}%
\newcommand\skytel{\refer@jnl{S\&T}}%
\newcommand\solphys{\refer@jnl{Sol.~Phys.}}%
\newcommand\sovast{\refer@jnl{Soviet~Ast.}}%
\newcommand\ssr{\refer@jnl{Space~Sci.~Rev.}}%
\newcommand\zap{\refer@jnl{ZAp}}%
\newcommand\nat{\refer@jnl{Nature}}%
\newcommand\iaucirc{\refer@jnl{IAU~Circ.}}%
\newcommand\aplett{\refer@jnl{Astrophys.~Lett.}}%
\newcommand\apspr{\refer@jnl{Astrophys.~Space~Phys.~Res.}}%
\newcommand\bain{\refer@jnl{Bull.~Astron.~Inst.~Netherlands}}%
\newcommand\fcp{\refer@jnl{Fund.~Cosmic~Phys.}}%
\newcommand\gca{\refer@jnl{Geochim.~Cosmochim.~Acta}}%
\newcommand\grl{\refer@jnl{Geophys.~Res.~Lett.}}%
\newcommand\jcp{\refer@jnl{J.~Chem.~Phys.}}%
\newcommand\jgr{\refer@jnl{J.~Geophys.~Res.}}%
\newcommand\jqsrt{\refer@jnl{J.~Quant.~Spec.~Radiat.~Transf.}}%
\newcommand\memsai{\refer@jnl{Mem.~Soc.~Astron.~Italiana}}%
\newcommand\nphysa{\refer@jnl{Nucl.~Phys.~A}}%
\newcommand\physrep{\refer@jnl{Phys.~Rep.}}%
\newcommand\physscr{\refer@jnl{Phys.~Scr}}%
\newcommand\planss{\refer@jnl{Planet.~Space~Sci.}}%
\newcommand\procspie{\refer@jnl{Proc.~SPIE}}%

\layoutstyle{6x9}

\newcommand{\iso}[2]{\ensuremath{\mathrm{^{#1}#2}}}

\newcommand{\mdot}{\ensuremath{\dot{m}}}
\newcommand{\medd}{\ensuremath{\mdot_{\mathrm{Edd}}}}

\newcommand{\vsed}{\ensuremath{w_{\mathrm{sed}}}}

\newcommand{\CP}{\ensuremath{C_P}}
\newcommand{\enuc}{\ensuremath{\varepsilon_\mathrm{nucl}}}
\newcommand{\ecool}{\ensuremath{\varepsilon_\mathrm{cool}}}
\newcommand{\gpscps}{\ensuremath{\gram\nsp\cm^{-2}\nsp\second^{-1}}}

\newcommand{\ee}[1]{\ensuremath{\times 10^{#1}}}
\newcommand{\yturn}{\ensuremath{y_{\mathrm{turn}}}}

\begin{document}
\title{Type I X-ray Bursts at Low Accretion Rates}


\author{Fang Peng}
{address={Theoretical Astrophysics, California Institute of Technology, Pasadena, CA 91125}}

\author{Edward F. Brown}
{address={Department of Physics \& Astronomy, Michigan State University, East Lansing, MI 48824}}

\author{James W. Truran}
{address={Department of Astronomy \& Astrophysics, University of Chicago, Chicago, IL 60637}, altaddress={Physics Division, Argonne National Laboratory,  Argonne, IL 60439}}

\begin{abstract}
Neutron stars, with their strong surface gravity, have interestingly short timescales for the sedimentation of heavy elements. Recent observations of unstable thermonuclear burning (observed as X-ray bursts) on the surfaces of slowly accreting neutron stars ($< 0.01$ of the Eddington rate) motivate us to examine how sedimentation of CNO isotopes affects the ignition of these bursts. We further estimate the burst development using a simple one-zone model with a full reaction network. We report a region of mass accretion rates for weak H flashes. Such flashes can lead to a large reservoir of He, the unstable burning of which may explain some observed long bursts (duration $\sim 1000$ s).
\end{abstract}

\maketitle

\section{Introduction}\label{introduction}

An ionized plasma in a gravitational field develops an electric field sufficient to levitate the ions and ensure overall charge neutrality. When there is more than one species of ion present, the ions will experience a differential force: lighter ions float upward (defined by the local gravitational field) and heavier ions sink downward. 

Accreting neutron stars, with their strong surface gravity $\approx 2.0\times
10^{14}\nsp\cm\usp\second^{-2}$, are an ideal place to look for the effects of sedimentation.  The sedimentation of heavy elements and resulting nucleosynthesis in the envelope of isolated neutron stars cooling from birth was first described by \citet{rosen:nucleosynthesis,rosen:particle} and has been studied in detail by \citet{chang.bildsten:diffusive,chang.bildsten:evolution}. For accreting neutron stars, the rapid stratification removes heavy nuclei from the photosphere for accretion rates $\dot{M} \lesssim 10^{-12}\Msun\usp\yr^{-1}$ \citep*{bildsten92}. Deeper in the neutron star envelope, the differentiation of the isotopes can alter the nuclear burning, namely unstable H/He burning and rapid proton-capture process (rp-process), that powers Type I X-ray bursts. 

Recent long-term monitoring of the galactic center with \beppo\ led to the discovery of nine ``burst-only sources'' \citep[see][and references therein]{Cornelisse2004Burst-only-sour}. These sources did not have persistent fluxes detectable with the \beppo/WFC, thus must have extremely low accretion luminosities ($L_X < 10^{36}\usp\ergspersecond \simeq 0.01\ensuremath{L_{\mathrm{Edd}}}$).  If the accretion is not concentrated onto a small surface area, so that the local accretion rate is $\mdot\approx \dot{M}/(4\pi R^{2})$, then the sedimentation timescale for CNO nuclei, defined as the time required to move a scale height relative to the center of mass of a fluid element, is less than the accretion timescale, defined as the time for a fluid element to reach a given depth. An estimate of this accretion rate was noted earlier by \citet{wallace.woosley.ea:thermonuclear}. As a result, sedimentation in the accreted envelope must be considered in treating the unstable ignition of hydrogen and helium for these low-$\mdot$ sources. At somewhat higher accretion rates, $L_X \sim 0.01L_{\mathrm{Edd}}$, several bursts have been observed with longer durations, $\lesssim 1000\usp\second$, that are intermediate between mixed H/He bursts and superbursts. 

Motivated by these observations, we explore the unstable ignition of hydrogen and helium at low accretion rates and pay particular attention to the regime where the mass accretion rate is less than the critical rate needed for stable H burning \citep*{fujimoto81:_shell_x}. We also study how the sedimentation of CNO nuclei affects the unstable ignition of H and He in an accreted neutron star envelope. For a full reporting of these results, please see \citep{Peng2006Sedimentation-a}.

\section{The Role of Sedimentation on Ignition Conditions}

To study the accumulation of H and He, we construct models of a plane-parallel atmosphere is in hydrostatic equilibrium, and for simplicity we neglect the terms coupling thermal and particle diffusion \citep[see][and references therein]{paquette.pelletier.ea:diffusion}.  We follow the treatment of  \citet{burgers69:composite_gases} and construct for each species an equation of continuity and momentum conservation. Collisions are described by a resistance coefficient $K_{st} = n_s n_t \langle \sigma_{st} v_{st} \rangle$, with $v_{st}$ the center-of-mass relative velocity between particles of types $s$ and $t$.

As an illustration of how sedimentation changes the structure of the envelope, consider a trace species, labeled 2, in a background of a species labeled 1, i.e., $n_1 \gg
n_2$ and species 1 has velocity $w_1 = 0$.  The electric field is $eE = \mb g A_1/(Z_1 + 1)$
where the electrons are non-degenerate and $eE = \mb g A_1/Z_1$ where
the electrons are degenerate.  Substituting $E$ into the
equation of motion for species 2 then determines the sedimentation velocity,
$\vsed = w_2 = n_2 (A_2 \mb g - Z_2 eE)/K_{12}$.  We chose the
sign of $\vsed$ to be positive if species 2 moves downward.

Using the Stokes-Einstein relation to determine $K_{12}$ from the drag
coefficient for a liquid sphere (\cite{bildsten.hall:diffusion}), and a nonrelativistic electron equation of state, we find the sedimentation velocity of a trace nucleus 
\citep*[see][]{brown.bildsten.ea:variability} 
\begin{equation}\label{e.vsed}
\vsed = 2 \times 10^{-3} g_{14}\frac{T_7^{0.3}}{\rho_5^{0.6}}
\frac{(A_2Z_1 - Z_2A_1)A_1^{0.1}}{Z_1^{2.3}Z_2^{0.3}}\nsp\cm\usp\second^{-1}, 
\end{equation}
where we use the common shorthand 
$g_{14} = g/(10^{14}\nsp\cm\usp\second^{-2})$, $T_7 = T/(10^7\usp\K)$, and $\rho_5 = \rho/(10^5\usp\grampercc)$.
In a pure H plasma, the sedimentation velocity\footnote{Throughout this paper we use a
  Newtonian metric and assume a neutron star mass $1.4\nsp\Msun$ and
  radius 10\nsp\kilo\meter, so that $g_{14} = 1.9$.} is greater than the mean velocity $u = \mdot/\rho$ for 
$\mdot < 400 \usp\gpscps T_7^{0.3} \rho_5^{0.4} (A_2 - Z_2)/Z_2^{0.3}$; under
conditions at which H ignites ($T_7 \sim 5; \rho_5 \sim 4$) 
this corresponds to $\mdot < 0.02\medd$ and
$\mdot < 0.05 \medd$ for a trace \iso{4}He and
\iso{12}C nucleus, respectively. Here $\medd = 2 m_{\rm p} c/[(1+X_{\rm H}) \sigma_{\rm TH} R] = 8.8\times 10^{4}\nsp\gpscps$ is the local Eddington mass accretion rate for a solar composition,
$\sigma_{\rm TH}$ is the Thomson scattering cross section, and $X_{\rm H}$ is the hydrogen mass fraction. 

We numerically solve for the evolution of the accreting envelope prior to unstable ignition by coupling the diffusion equations to the equations for temperature and heat flux,
\begin{equation}\label{heat.e}
  \frac{\partial T}{\partial y} = \frac{F}{\rho K}\ , \qquad
  \frac{\partial F}{\partial y} = 
  \CP\left(\frac{\partial T}{\partial t} + 
    \mdot\frac{\partial{T}}{\partial{y}}\right)
  - \frac{\CP T\mdot}{y}\nabla_\mathrm{ad} - \enuc\ ,
\end{equation}
where $dy = - \rho dz$ is the column density, \CP\ is the specific heat at constant pressure and
$\nabla_\mathrm{ad} \equiv (\partial\ln T/\partial\ln P)_S$. 
For detailed description of the numerical method in solving these equations, please refer to 
\citep{Peng2006Sedimentation-a}. 

We take the composition of matter accreted from the companion star to be roughly solar composition in \iso{1}{H} and \iso{4}{He}, and distribute the remaining mass evenly between
\iso{12}{C} and \iso{16}{O}:
$X(\iso{1}{H})=0.71$, $X(\iso{4}{He})=0.27$, $X(\iso{12}{C}) = 0.01$, 
and $X(\iso{16}{O}) = 0.01$.  Below the fuel layer are ashes
from previous X-ray bursts; we set the ash composition to \iso{64}{Zn}, consistent with the findings of recent one-dimensional calculations of repeated X-ray bursts \citep{woosley.heger.ea:models,Fisker2005Extracting-the-}.  We use a radiative-zero outer boundary, and
for the inner boundary, at the bottom of the ash layer, we set the flux to a constant value, 
$F_{\rm c}=0.1\nsp\MeV \nsp\mdot/\mb$ \citep{brown:nuclear}. This flux is set by reactions in the deep crust---electron captures, neutron emissions, and pycnonuclear reactions.

\begin{figure}[t]
\centering
\includegraphics[width=0.6\textwidth]{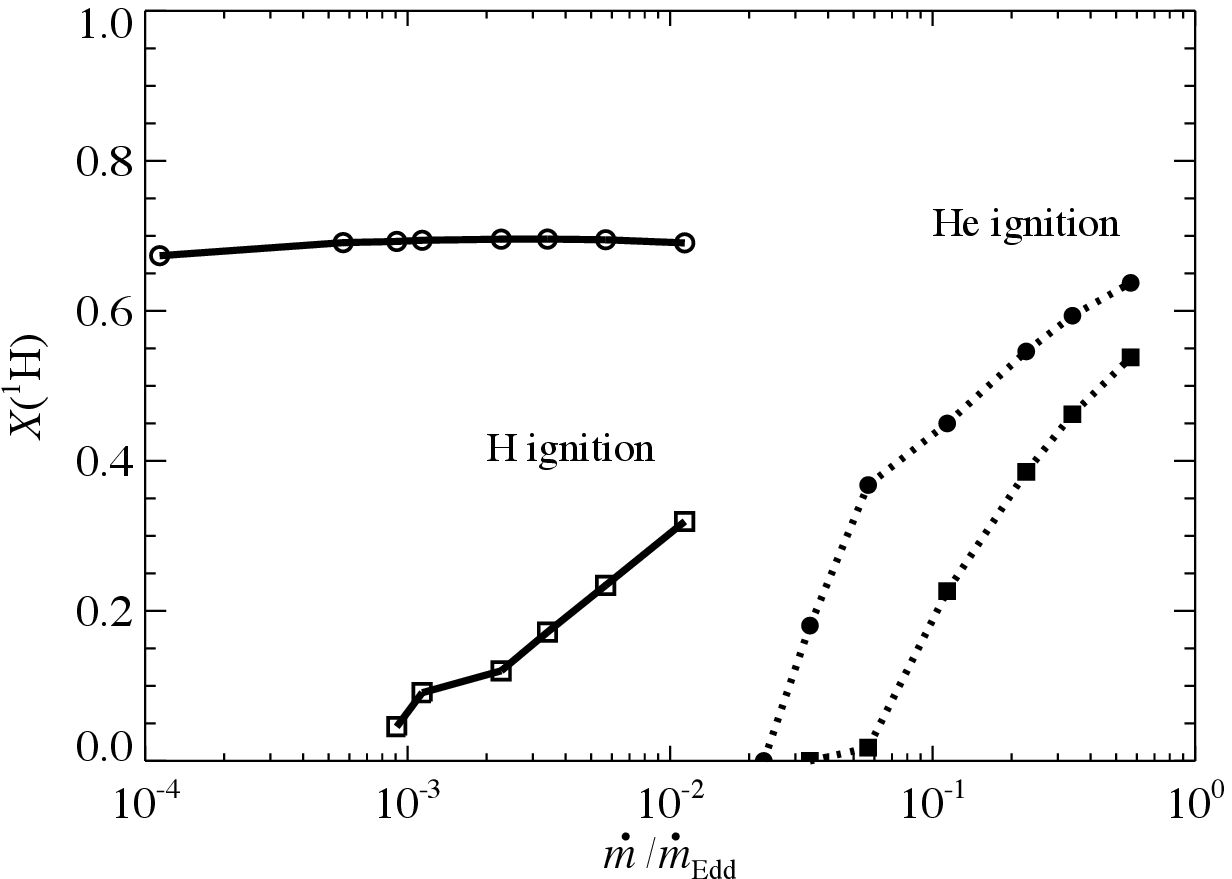}
\caption{
  Mass fraction of hydrogen at the column where either H (\emph{open symbols, solid lines}) or He
  (\emph{filled symbols, dotted lines}) unstably ignites, as a function of mass accretion rates.  We show results for which sedimentation is ignored (\emph{circles}), and for which it is included (\emph{squares}). From \citep{Peng2006Sedimentation-a} and reproduced by permission of the AAS.}
 \label{abund_mdot.f}
\end{figure}

We explore the evolution of the accreting neutron
star envelope for accretion rates $10^{-4}\textrm{--}0.6\nsp\medd$ ($10\textrm{--}5\times10^{4}\nsp\gpscps$ for accretion of a solar composition mixture).  For each accretion rate, we ran calculations both with and without isotopic sedimentation, and we follow the thermal and chemical evolution of the mixture until the envelope becomes thermally unstable. 
We determine this ignition point using a simple one-zone  criterion,
$(\partial\enuc/\partial T)_P > (\partial \ecool/\partial T)_P$, where 
\enuc\ is the nuclear heating rate and $\ecool = \rho KT/y^2$ is an 
approximation of the local cooling rate. The mass fraction of H and He at the column where either
H or He unstably ignites is plotted in fig.~\ref{abund_mdot.f}, 
as a function of mass 
accretion rates. When sedimentation is included, the abundance of H at the base of the 
accreted envelope is depressed. It is tantalizing that the accretion rate at which mixed H/He ignition occurs is increased by a factor of 2 when sedimentation is taken into account, and we speculate that this might alleviate the discrepancy between the predicted transition in burst duration \citep{fujimoto81:_shell_x} and recent observations \citep[see, for example][]{den-Hartog2003Burst-propertie}

\section{Bursts at Low Accretion Rates}\label{sec:bursts-at-low}
When the temperature in the neutron star envelope is sufficiently low, the CNO cycle becomes temperature dependent; as a result, the ignition of H becomes thermally unstable at low accretion rates \citep{fujimoto81:_shell_x}. As a first investigation of the unstable ignition of H when the atmosphere is stratified, we perform several calculations of the burst nucleosynthesis.  We approximate the cooling by a one-zone finite differencing over the envelope,
\begin{equation}
  \CP\frac{dT}{dt} = \enuc - \ecool,   \label{eq:one-zone}
\end{equation}
where $\ecool = \rho K T y^{-2}$ and we evaluate \enuc\ from a reaction network.
Included in this network are 686 isotopes covering proton-rich nuclei up to Xe \citep[see][]{schatz.aprahamian.ea:endpoint}. The reaction rates are taken from the compilation \code{reaclib} \citep[see][and references therein]{thielemann86,rauscher00}, and consist of experimental rates and  Hauser-Feshbach calculations with the code \nonsmoker\ \citep{rauscher00}.  The initial temperature, column density, and composition are taken from the values of the quasi-static calculation at the ignition point. For the unstable ignition of H, the temperature in the accreted envelope is set by the flux emergent from the deeper ocean and crust.  As a result, the accretion rate at which the burst behavior changes will depend on assumptions about the heating in the crust. 

\begin{figure}[bt]
\centering
\includegraphics[width=0.49 \textwidth,viewport=0 0 290 310,clip]{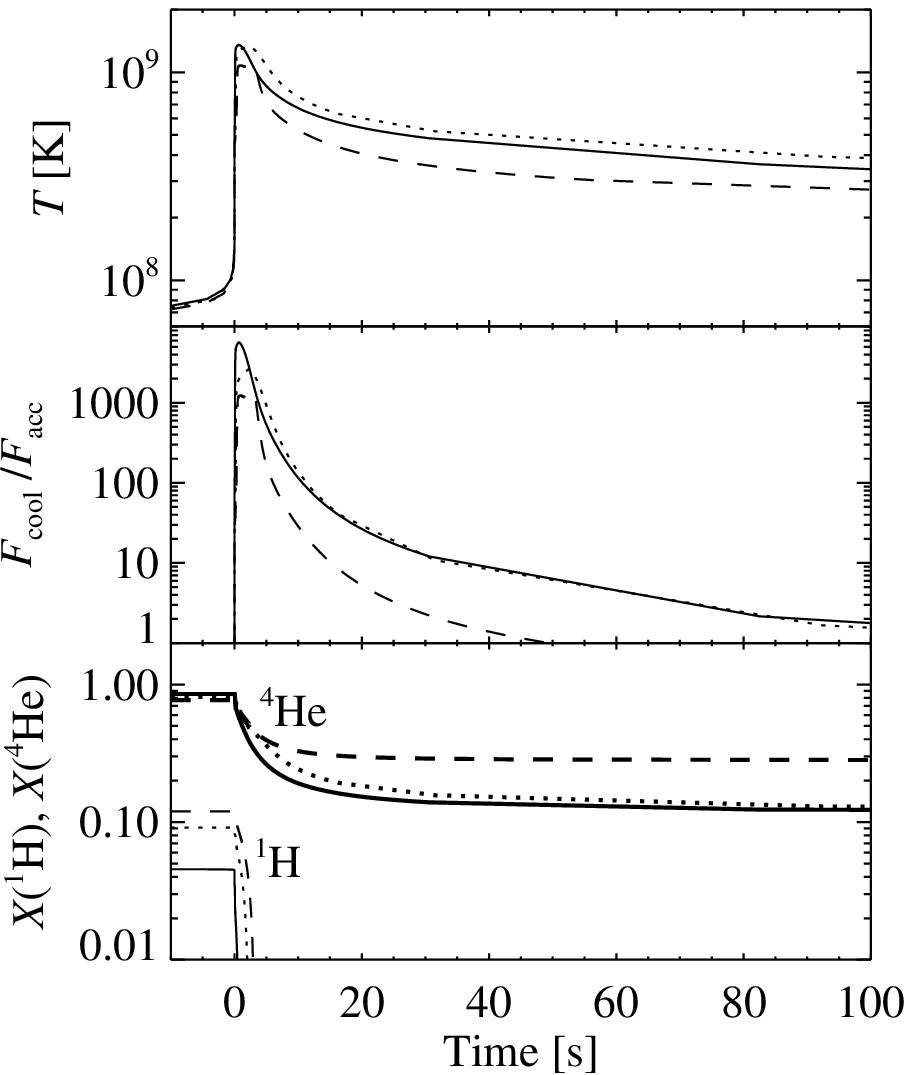}
\includegraphics[width=0.49 \textwidth,viewport=0 0 290 310,clip]{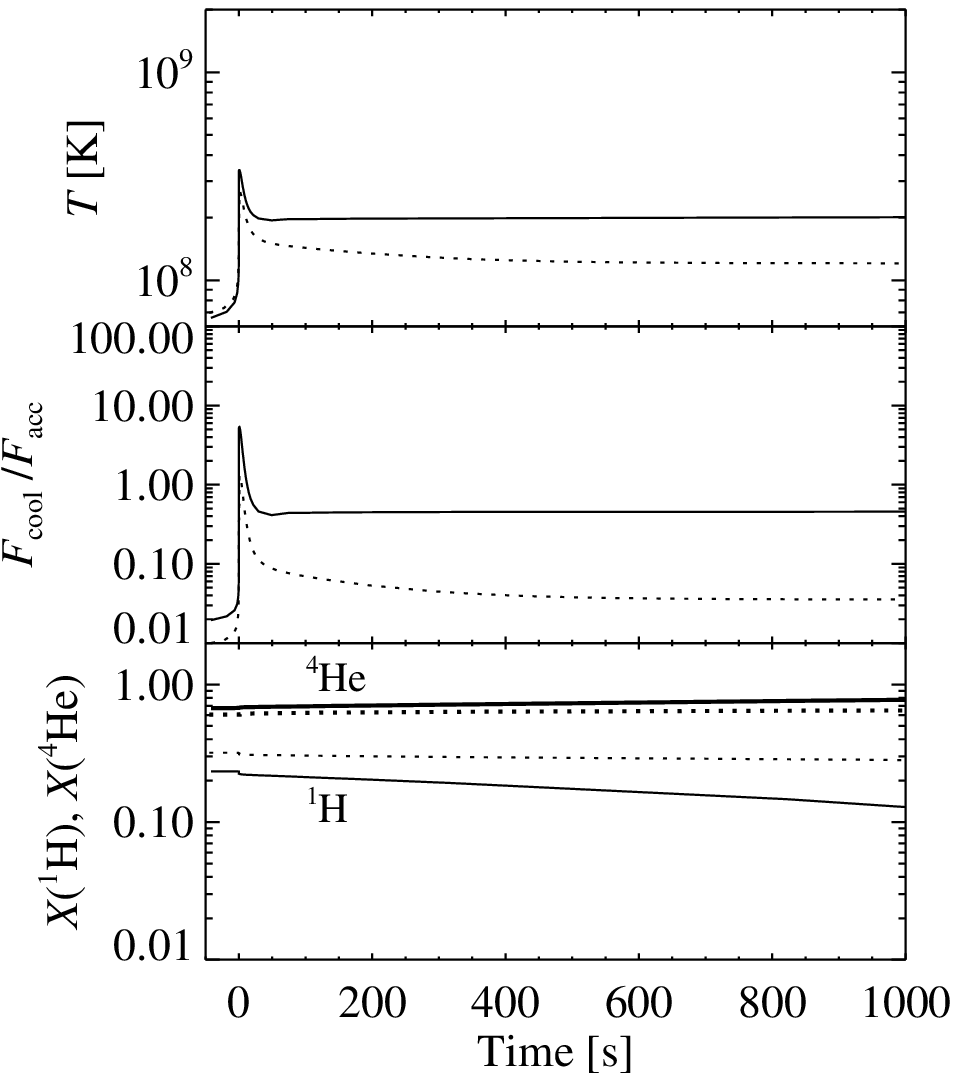}
\caption{One-zone burst calculation following unstable H ignition. Left: for three mass accretion rates, $\mdot/\medd = 9.1\ee{-4}$ (\emph{solid lines}), $1.1\ee{-3}$ (\emph{dotted lines}) and $2.3\ee{-3}$ (\emph{dashed lines}), respectively. Right: for two mass accretion rates, $\mdot/\medd = 5.7\ee{-3}$ and 0.011. Top panel: temperature evolution; Middle panel: the ratio of one-zone cooling flux to the accretion flux; Bottom panel: mass fraction of hydrogen (\emph{thin lines}) and $\iso{4}{He}$ (\emph{thick lines}), respectively. Diffusion and sedimentation is included. From \citep{Peng2006Sedimentation-a}, and reproduced by permission of the AAS.}
\label{bursts.f}
\end{figure}

We shall first describe the outcome of these one-zone calculations (eq.~[\ref{eq:one-zone}]). 
The left panel of Figure~\ref{bursts.f} shows the post-ignition evolution for accretion rates $9.1\ee{-4}\nsp\medd$ (80\nsp\gpscps; \emph{solid lines}),  $1.1\ee{-3}\nsp\medd$ (100\nsp\gpscps; \emph{dotted lines}), and $2.3\ee{-3}\nsp\medd$ (200\nsp\gpscps; \emph{dashed lines}). We plot three different quantities: the evolution of temperature (\emph{top panel}), the heat flux,
$F_{\rm cool} = y \ecool$, normalized to the accretion flux (\emph{middle panel}), and the mass fractions of hydrogen and helium (\emph{bottom panel}). We then repeat this calculation at higher mass accretion rates (Fig.~\ref{bursts.f}, right panel), $5.7\ee{-3}\nsp\medd$ (500\nsp\gpscps; \emph{solid lines}) and $1.1\ee{-2}\nsp\medd$ ($10^{3}\nsp\gpscps$; \emph{dotted lines}). 

It is immediately clear that there are two very different outcomes for the H ignition. At lower accretion rates, the rise in temperature following H ignition is sufficient to trigger a vigorous He flash with a decay timescale $\sim 10\nsp\second$ (Fig.~\ref{bursts.f}, \emph{middle-left panel}).  At higher accretion rates, however, the flash is too weak to ignite He. As an interpretation of these results, recall that the ignition curve for the $3\alpha$ reaction has a turning point at $\yturn\approx 5\ee{7}\nsp\GramPerSc$,  which terminates the unstable branch \citep{bildsten:thermonuclear,cumming.bildsten:rotational}.  At the higher accretion rates (Fig.~\ref{bursts.f}, right panel) the ignition of H occurs at $y < \yturn$, and the local rise in temperature does not trigger unstable He ignition. In fact, the temperature rise is not even sufficient to initiate convection, as the radiative temperature gradient needed to carry $F_{\mathrm{cool}}$ is $d\ln T/d\ln P \approx 0.25 < (\partial\ln T/\partial\ln P)_{s}$.  In contrast, at lower accretion rates (Fig.~\ref{bursts.f}, left panel), the ignition of H occurs at $y > \yturn$, and the rise in temperature will ignite the triple-$\alpha$ reaction.  

If subsequent H flashes do not ignite the underlying He, then a large layer of nearly pure He will accumulate.  Because our system is not burning H steadily, the temperature in the He layer is colder than if the burning were in steady-state. \emph{As a result,  a large He layer should accumulate.}
To get the thermal structure of an accumulated pure He layer, we integrated equations (\ref{heat.e}), with $\partial/\partial t\to 0$, from the column where H ignition occurred, and set the temperature there to that found in the accumulating model. 
For $\mdot/\medd=5.7\ee{-3}$ and 0.011, the He ignition column density varies from 
$y\approx 2\times10^{9}\textrm{--}3\times10^{11}\nsp\gram\usp\cm^{-2}$ (recurrence time $\approx 0.06-19\nsp\yr$) for $F_{\rm c} \mb/\mdot$ varies from $1.0$ to $0.1$. Our one-zone approximation gives a cooling time  $\tau \approx 400\nsp\second [y/(10^{10}\nsp\gram\usp\cm^{-2})]$, which roughly agrees with the long burst duration. 

\section{Conclusion}

Using a simplified numerical model of an accreting neutron star envelope that allows for differential isotopic velocities, we have undertaken a first study of the effects of sedimentation and diffusion on the unstable ignition of hydrogen and helium on the surfaces of accreting neutron stars, and have investigated the outcome of unstable hydrogen ignition using simple one-zone models. Our principal conclusions are:
\begin{enumerate}
\item The effect of sedimentation changes the conditions at H/He ignition even for $\mdot \gtrsim 0.1\medd$.

\item There is a range of accretion rates for which unstable H ignition does not trigger a He flash.  This range depends on the flux from the deep crustal heating and the degree of settling of He and CNO nuclei. We speculate that successive weak H flashes can lead to the accumulation of a large reservoir of He; this may explain the long bursts observed from some sources.
\end{enumerate}

\begin{theacknowledgments}
This work is supported in part by the \textbf{J}oint \textbf{I}nstitute for \textbf{N}uclear \textbf{A}strophysics under NSF-PFC grant PHY~02-16783, and by the Department of Energy under grant B523820 to the Center for Astrophysical Thermonuclear Flashes at the University of Chicago. EFB is supported by the NSF under grant AST-0507456; JWT acknowledges support from the U.S. DOE, under contract No. W-31-109-ENG-38.
\end{theacknowledgments}


\end{document}